\documentclass[%
 reprint,
superscriptaddress,
 amsmath,amssymb,
 aps,
prb,
]{revtex4-2}

\usepackage{graphicx}
\usepackage{dcolumn}
\usepackage{bm}
\usepackage{float}
\usepackage{graphicx} 
\newcommand{\vect}[1]{\boldsymbol{#1}}

\begin{document}




\preprint{APS/123-QED}

\title{Giant spin Hall angle in the Heusler alloy Weyl ferromagnet Co\textsubscript{2}MnGa}

\author{L. Leiva}
\affiliation{Department of Electronic Science and Engineering, Kyoto University, Kyoto 615-8510, Japan}

\author{S. Granville}
\author{Y. Zhang}
\affiliation{Robinson Research Institute, Victoria University of Wellington, Wellington 6140, New Zealand}
\affiliation{The MacDiarmid Institute for Advanced Materials and Nanotechnology, Wellington 6011, New Zealand}
\author{S. Dushenko}
\affiliation{Department of Electronic Science and Engineering, Kyoto University, Kyoto 615-8510, Japan}
\affiliation{Institute for Research in Electronics and Applied Physics, University of Maryland, College Park, MD 20742, USA}
\affiliation{Physical Measurements Laboratory, National Institute of Standards and Technology, Gaithersburg, MD 20899, USA}
\author{E. Shigematsu}
\author{T. Shinjo}
\author{R. Ohshima}
\author{Y. Ando}
\author{M. Shiraishi}
\affiliation{Department of Electronic Science and Engineering, Kyoto University, Kyoto 615-8510, Japan}

\date{\today}

\begin{abstract}
Weyl semimetals are playing a major role in condensed matter physics due to exotic topological properties, and their coexistence with ferromagnetism may lead to enhanced spin-related phenomena. Here, the inverse spin Hall effect (ISHE) in the ferromagnetic Weyl-semimetal Heusler alloy Co$_2$MnGa was investigated at room temperature by means of electrical spin injection in lateral spin valve structures. Spin transport properties such as spin polarization and spin diffusion length in this material were precisely extracted in order to estimate the spin Hall angle $\theta_{\textrm{SH}}$, which was found to be $-0.19\pm0.04$ and is among the highest reported for a ferromagnet. Although this value is on the same order of magnitude of known heavy metals, the significantly higher resistivity of Co$_2$MnGa implies an improvement on the magnitude of detection voltages, while its ferromagnetic nature allows controlling the intensity of SHE through the magnetization direction. It was also shown that Onsager's reciprocity does not hold for this system, which is in part attributable to a different spin-dependent Hall conductivity for spin-up and spin-down carriers.
\end{abstract}

\maketitle

Weyl semimetals have been attracting significant attention since the discovery of a nonmagnetic Weyl material, TaAs \cite{TaAs}, because of its band-crossing points that give rise to plenty of unique physical properties, such as the Fermi arc surface states, the chiral anomaly coming from Nielsen-Ninomiya theorem \cite{Nielsen} and monopole-like Berry curvature \cite{TopologicalWeyl}. Furthermore, a prominent class of Weyl semimetals is Weyl magnetic materials such as the ferromagnetic Heusler alloy Co$_2$MnGa \cite{Sakai,Weyl} and antiferromagnetic Mn$_3$Sn \cite{Mn3SnWeyl}. These two materials are playing pivotal roles in condensed-matter physics because of the recent discoveries of gigantic anomalous Hall effect (AHE) \cite{Simon1,AHEMn3Sn}, magnetic spin Hall effect (a novel family of Hall effects) \cite{MagneticSHE} and spin caloritronics phenomena such as a large anomalous Nernst effect (ANE) \cite{Sakai,reichlova2018large}. Additionally, magnetic Heusler alloys have emerged as promising materials in the field of spintronics due to their either half-metallic or semimetallic nature, which would lead to a high spin polarization \cite{Coey,Spintronics}, as has been reported in Co-based full Heusler compounds \cite{KimuraIntro,Shirotori}. Combining the remarkable spin transport properties of Heusler alloys with the unique band structure of Weyl semimetals may lead to new exotic phenomena for topologically driven spintronic applications.

In the quest for these novel phenomena, an old acquaintance has emerged, the spin Hall effect (SHE). Together with its reciprocal version, the inverse spin Hall effect (ISHE), these two are utilized as essential methods for the generation and detection of pure spin currents in spintronics devices, since they enable the conversion of a charge current into a transversal spin current and vice versa \cite{Sinova}. These spin-orbit coupling phenomena have been widely observed in many non magnetic heavy metals (HMs) \cite{Saitoh,HeavyMetalSH,tao2018self,Morota,Mosendz}, but it is thought that replacing a HM with a ferromagnet (FM) offers potential advantages such as precise control of the spin current through the magnetization direction, which could be applied to spin-transfer torque devices \cite{Taniguchi}. However, measurements on only a few ferromagnetic materials have been reported to date, most of them exhibiting spin Hall angles of just a few percentage points \cite{Tsukahara,SHEPy,SHEFePt,Seki,Oomori,Iihama,Wimmer}. To overcome this limitation, materials exhibiting a large AHE are required, since it is understood that both the AHE and the SHE are driven by the same intrinsic and extrinsic scattering mechanisms, related to the spin-orbit interaction \cite{Zimm,Hoffmann}. The large Berry curvature distribution around the Fermi level in Co$_2$MnGa, thought to be responsible for its large AHE \cite{Markou}, and \textit{ab initio} calculations that predict a strong intrinsic spin Hall effect in other Weyl semimetals \cite{Sun}, indicate Co$_2$MnGa is a strong candidate to observe large spin Hall voltages. In addition, significant spin polarization \cite{Kolbe,Tung_2013}, strong resistance to oxidation \cite{Simon1} and a higher resistivity than conventional metals \cite{Holmes} suggest that Co$_2$MnGa is a suitable platform to study the spin transport by means of electrical spin injection. In this work, the ISHE of Co$_2$MnGa is investigated, wherein a giant spin Hall angle of $-0.19\pm0.04$ is found, which is among the highest reported for a FM so far.

\begin{figure}
	\includegraphics[width=8.6cm]{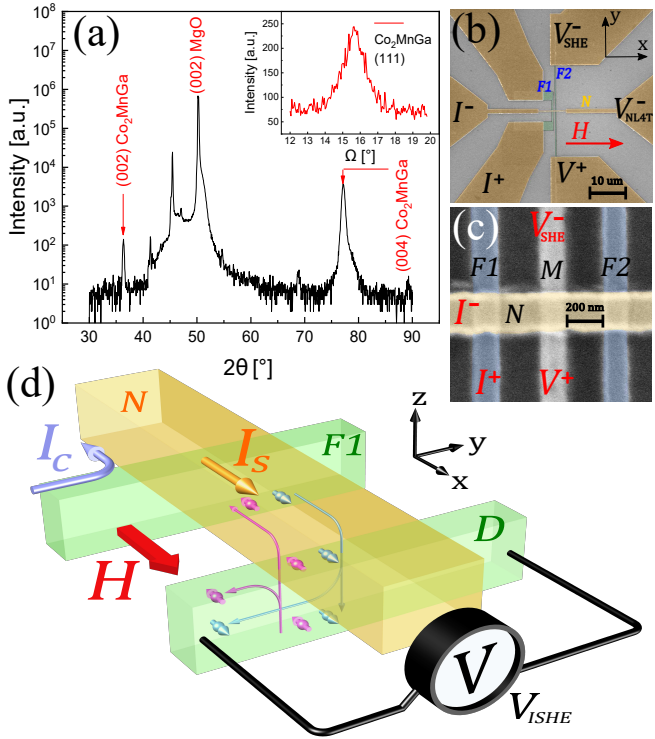}
	\centering
	\caption{(a) XRD pattern obtained for an out-of-plane $\theta$-2$\theta$ scan of a sample consisting of a 30 nm-thick Co$_2$MnGa film grown on a single-crystal MgO substrate. Only the reflections due to the main X-ray source (Co-K$_\alpha$) are indexed. Inset shows the rocking curve for the (111) peak. (b) Scanning electron microscope (SEM) image of a 2-wire LSV. In an ISHE/SHE measurement, the magnetic field $H$ is applied parallel to the copper channel (Colored in edition). (c) SEM image of a close view to the electrodes in a LSV with an absorption middle wire (Colored in edition). (d) Schematics of the nonlocal ISHE measurement. A charge current $I_c$ is injected from $F1$ to $N$, producing spin accumulation at the $F1/N$ interface and a consequent spin current $I_s$ diffuses along $N$. Part of this spin current is absorbed by the detector $D$ in negative $z$ direction. Then, a charge current in positive $y$ direction is induced in $D$ due to the inverse spin Hall effect, and a voltage can be measured in equilibrium open circuit condition.}
	\label{Uno}
\end{figure}

\begin{figure}
	\includegraphics[width=8cm]{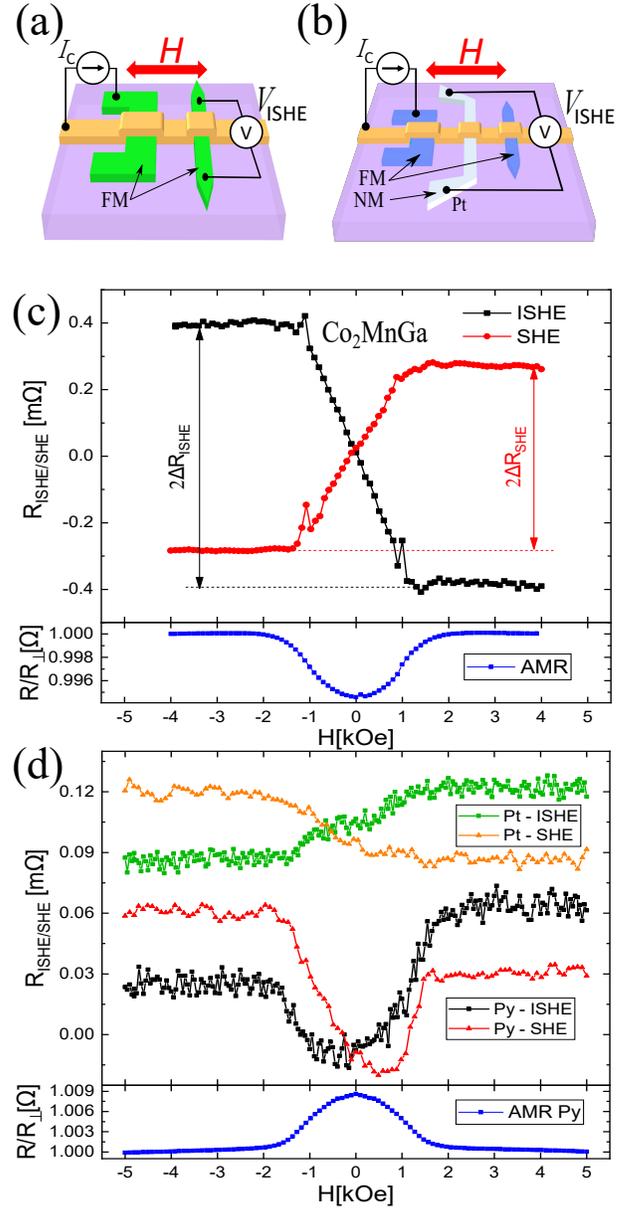}
	\centering
	\caption{(a) Schematics of ISHE setup for a ferromagnetic detector. In this kind of samples both FM electrodes are made of the same material, either Co$_2$MnGa or Py. For SHE setup, current source and voltmeter are exchanged preserving the polarity. (b) Schematics of ISHE setup for a non-magnetic detector. In this case Py was used for FM electrodes, but one of them is not used in this measurement. SHE setup is obtained by exchanging the current source with the voltmeter while preserving the polarity. (c) Nonlocal resistance for the ISHE setup and for its reciprocal (direct) SHE setup for Co$_2$MnGa, measured as a function of in-plane external magnetic field $H$ at room temperature. Bottom panel shows a typical AMR curve for Co$_2$MnGa electrode $F1$. (d) Equivalent nonlocal ISHE and SHE resistances measured for Pt and Py at room temperature. Bottom panel shows the AMR curve for the electrode $F1$ made of Py in both cases.}
	\label{Dos}
\end{figure}

The samples consisted of lateral spin valve (LSV) structures that were fabricated starting from 30-nm-thick films of Co$_2$MnGa epitaxially grown on a (001)-oriented MgO single-crystal substrate. An X-ray diffraction (XRD) $\theta$-2$\theta$ (out-of-plane) scan was performed to check the crystal structure in the films. The presence of (002) and (004) peaks of Co$_2$MnGa confirms $B2$ ordering in the thin films, as shown in Fig. \ref{Uno}(a), meanwhile the presence of (111) peak, as verified by the rocking curve in the inset, confirms the $L2_1$ ordering in the samples. The FM electrodes were patterned on the sub-micron scale using conventional electron-beam lithography and Ar ion milling techniques. Two types of devices were fabricated, one with two parallel FM electrodes (Fig. \ref{Uno}(b)) with various channel lengths $L$ (edge-to-edge separation) while the other with three electrodes (Fig. \ref{Uno}(c)) with a fixed distance of 600 nm between $F1$ and $F2$ and various widths of the middle wire $M$, in order to estimate the spin resistance through the absorption technique \cite{Morota,Oomori,laczkowski,Niimi1,Niimi2,sagasta2017spin,zahnd2018spin}. Finally, copper channels perpendicularly connecting the ferromagnetic electrodes, as well as the macroscopic connection pads, were patterned using electron-beam lithography and thermal evaporation techniques. It is worth mentioning that, in order to obtain Ohmic transparent interfaces, a low-acceleration-voltage \textit{in-situ} Ar ion milling was performed prior to copper deposition. Cu was selected for the non-magnetic channel because of its long spin diffusion length and long spin relaxation time \cite{Jedema}, which make it a typical material for nonlocal signal measurements. For reference, devices with the same geometry but with Py ferromagnetic electrodes were fabricated on a thermally oxidized Si substrate. In some of the 2-wire Py devices, an absorption middle wire of Pt was deposited as a reference non-magnetic material, as shown in the scanning electron microscope image of Fig. \ref{Uno}(c). All the transport measurements were performed at room temperature in a commercial physical property measurement system (PPMS) using a DC technique that consists of averaging the absolute value of the voltage measured with positive and negative DC currents, which is equivalent to an AC lock-in technique \cite{Casanova}. Further details of the fabrication procedure, as well as testing of the transparent interfaces and the determination of the spin diffusion length of Co$_2$MnGa are presented in the Supplemental Material in Secs. A, B and C \cite{SuppMat}.

Figure \ref{Uno}(d) shows the scheme of a typical ISHE measurement setup in a LSV. An electric current $I_c$ is injected from the $F1$ electrode into the left side of the $N$ channel (terminals $I^+$ and $I^-$ in Figs. \ref{Uno}(b) and \ref{Uno}(c)) producing spin accumulation at the $F1/N$ interface that induces a diffusive pure spin current along the $N$ wire \cite{Valenzuela}. Part of the spin current is then absorbed vertically (negative $\vect{z}$ direction) in a detection electrode $D$ ($D$ can be $F2$ or $M$ depending on the device). Since the spin orientation $\vect{\hat s}$ of the conduction electrons is given by the magnetization of $F1$, which is fixed by the external magnetic field applied along the $N$ channel, a charge current density $\vect{j_c}$ given by $\vect{j_c}=\theta_{SH}\vect{\hat s}\times \vect{j_s}$ \cite{SpinHallEffectDef} (where $\theta_{SH}$ is the spin Hall angle and $\vect{j_s}$ is the spin current density) is generated along the $D$ electrode length. In the equilibrium state in open circuit condition, a voltage $V_{\textrm{ISHE}}$ is generated to suppress the charge current in $D$, measured between terminals $V^+$ and $V^-_\textrm{SHE}$. When the magnetization of the injection electrode is switched, by sweeping the external magnetic field, the sign of the generated voltage is also expected to switch, as it is shown in the ISHE labeled measurements in Fig. \ref{Dos}(c) for Co$_2$MnGa and Fig. \ref{Dos}(d) for Pt and Py as detector electrodes, where $R_{\textrm{ISHE}}=V_{\textrm{ISHE}}/I_c$. The difference between the saturation values of $R_{\textrm{ISHE}}$ for positive and negative field is defined as $2\Delta R_{\textrm{ISHE}}$. Reciprocally, if the probe setup is inverted, by exchanging $V^+$ with $I^+$ and $V^-_\textrm{SHE}$ with $I^-$, a charge current along $D$ will inject a pure spin current in the $D/N$ interface by means of the (direct) SHE. This spin current will diffuse again across the $N$ channel to be detected as a nonlocal voltage $V_{\textrm{SHE}}$ between the electrodes $F1$ and $N$. The SHE setup nonlocal resistance $R_{\textrm{SHE}}$ (defined as $V_{\textrm{SHE}}/I_c$) is also shown in the SHE labeled measurements on Fig. \ref{Dos}(c) for Co$_2$MnGa and Fig. \ref{Dos}(d) for Pt and Py, where $2\Delta R_{\textrm{SHE}}$ can be compared with $2\Delta R_{\textrm{ISHE}}$. Note that the magnitudes of ISHE and SHE signals for Co$_2$MnGa are not the same, which is discussed later, and they are substantially large, roughly 20 times greater than for Pt and Py in very similar geometries. Anisotropic magnetoresistance (AMR) curves for the injection electrode $F1$ are shown to prove that the saturation of the ISHE and SHE signals correspond to the saturation of the magnetization of $F1$, where it is worth noting that the AMR of Co$_2$MnGa was found to be negative, as previously reported \cite{NegativeAMRCMG}.

For the ISHE measurement setup, the ISHE resistance $\Delta R_{\textrm{ISHE}}$ can be expressed as \cite{SpinHallEffectDef,Niimi1}:

\begin{equation}
\Delta R_{\textrm{ISHE}}=\frac{\langle I_s \rangle}{I_c}\theta_{SH}\frac{x}{W_D},
\end{equation}

\noindent where $\langle I_s \rangle$ is the spatial average of the absorbed spin current along the $\vect{z}$ direction, $W_D$ is the width of the detector wire and $x$ is the shunting factor that takes into account that part of the generated charge current is being shunted by the copper wire. By applying a one dimensional spin diffusion model \cite{1DSpin} for transparent interfaces and considering $t_D>>\lambda_D$, where $t_D$ the thickness of the detector electrode and $\lambda_D$ its spin diffusion length, $\Delta R_{\textrm{ISHE}}$ can be written as:

\begin{equation}
\Delta R_{\textrm{ISHE}}^{2W}=\frac{\theta_{SH} x W_N}{t_{F}}\frac{(1-\alpha_{F}^2)2\alpha_{F}R_F^2R_Ne^{-L/\lambda_N}}{(2R_{F}+R_N)^2-R_N^2e^{-2L/\lambda_N}}
\label{Rishe2W}
\end{equation}

\noindent for a 2-wire LSV using $F2$ as detector and

\begin{eqnarray}
\label{Rishe3W}
\Delta R_{\textrm{ISHE}}^{3W}&=&\frac{\theta_{SH} x W_N(1-\alpha_{M}^2)}{t_{M}}\\&\times&
\text{\footnotesize $\frac{2\alpha_{F}R_FR_MR_Ne^{-L/2\lambda_N}}{(2R_{F}+R_N)(2R_M+R_N)-(2R_M-R_N)R_Ne^{-L/\lambda_N}}$}\nonumber
\end{eqnarray}

\noindent for a 3-wire LSV using $M$ as detector, where $W_N$  is the width of $N$ wire, $\lambda_N$ is the spin diffusion length of Cu, $\alpha_F$ and $\alpha_M$ are the spin polarizations of  the ferromagnetic electrodes and the middle wire $M$ material, respectively, and $R_N$, $R_F$ and $R_M$ are respectively the spin resistances of $N$, $F1$ (considered to be the same as $F2$) and $M$ wires, defined in Ref. \cite{SuppMat}. For Eq. \ref{Rishe3W} it was considered that the middle wire $M$ was located at the middle of the gap distance $L$ between $F1$ and $F2$ electrodes.

\begin{figure}
	\includegraphics[width=8cm]{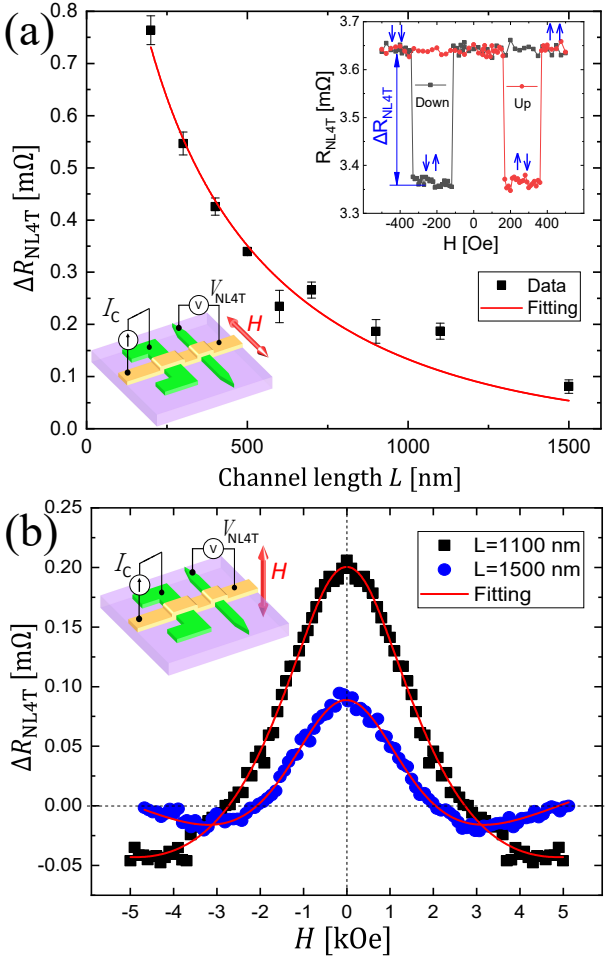}
	\centering
	\caption{(a) Channel length dependence of the nonlocal signal measured for several LSV devices at room temperature. The inset shows how is $\Delta R_{\textrm{NL4T}}$ determined in an example of the acquired spin signal as a function of in-plane longitudinal magnetic field. (b) Hanle effect nonlocal signal as a function of the external out-of-plane magnetic field, for two different channel lengths. A simultaneous fitting with common parameters is shown by the solid red line. In both figures, insets show a schematic of the measurement setup in each case.}
	\label{Tres}
\end{figure}

In the case when a FM material is used as a detector electrode, $\Delta R_{\textrm{ISHE}}$ can be written in terms of the conventional nonlocal-4-terminal (NL4T) resistance $\Delta R_{\textrm{NL4T}}$, which can be measured by connecting $V^-_\textrm{NL4T}$ terminal instead of $V^-_\textrm{SHE}$ in Fig. \ref{Uno}(b) and sweeping the external magnetic field in a direction parallel to the ferromagnetic electrodes length. A typical NL4T signal is shown in the inset of Fig. \ref{Tres}(a), where two different voltage levels are measured depending on whether the relative configuration of the magnetization of the ferromagnetic electrodes is parallel or anti-parallel. The difference between these two voltages, normalized by the injection current, define $\Delta R_{\textrm{NL4T}}$ that is a direct measure of the amount of spin current being absorbed by the detection wire. Then, Eqs. \ref{Rishe2W} and \ref{Rishe3W} are simplified into:

\begin{equation}
\Delta R_{\textrm{ISHE}}=\frac{\theta_{SH} x W_N}{t_{D}}\frac{(1-\alpha_{D}^2)}{2\alpha_{D}}\Delta R_{\textrm{NL4T}},
\label{RisheNL4T}
\end{equation}

\noindent where the only unknown spin transport parameter is the spin polarization of the detector electrode $\alpha_D$. 
In order to determine $\alpha_D$, $\Delta R_{\textrm{NL4T}}$ was measured in 2-wire LSVs for several different channel lengths $L$. This gap dependence is shown in Fig.\ref{Tres}(a) where the data has been fitted according to the equation \cite{1DSpin}:

\begin{equation}
\Delta R_{\textrm{NL4T}}=\frac{4\alpha_F^2R_F^2R_Ne^{-L/\lambda_N}}{(2R_{F}+R_N)^2-R_N^2e^{-2L/\lambda_N}},
\label{Rnl4t}
\end{equation}

\noindent where the only three unknown parameters are $\alpha_F$, $\lambda_N$ and $\lambda_F$, the spin diffusion length of the FM. Since $\alpha_F$ and $\lambda_F$ can not be extracted independently, additional spin absorption measurements were performed to determine $\lambda_F$ self-consistently, yielding  $\lambda_F=(3.1\pm1.1)$ nm for Co$_2$MnGa \cite{SuppMat}. The spin polarization $\alpha_F$ was also determined via Hanle effect nonlocal measurements, which have the same geometry and sample structure as in NL4T setup except for an out-of-plane external magnetic field instead of in-plane. The data for two different channel lengths is shown in Fig.\ref{Tres}(b), where the fitting was performed according to Eqs. (1), (2) and (3) in Ref. \cite{rojas2013plane} for the case of transparent interfaces, to consistently obtain $\alpha_F=0.15\pm0.03$.

\begin{figure}
	\includegraphics[width=8.6cm]{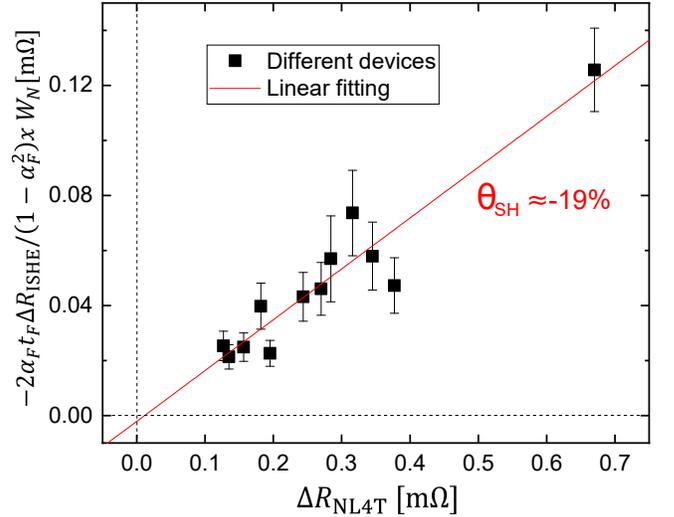}
	\centering
	\caption{Data of $-2\alpha_F t_F \Delta R_{\textrm{ISHE}}/(1-\alpha_F^2)W_N x$ as a function of $\Delta R_{\textrm{NL4T}}$ obtained for Co$_2$MnGa at room temperature for several devices. The spin Hall angle can be obtained from the slope of the linear fitting. Data points correspond to devices with different channel lengths $L$.}
	\label{Cuatro}
\end{figure}

The only unknown parameter left to estimate $\theta_{\textrm{SH}}$ is the shunting factor $x$, which was determined experimentally with devices specially designed to that end, as detailed in Sec. D of the Supplemental Material \cite{SuppMat}. Finally, a linear fitting, shown in Fig. \ref{Cuatro}, can be performed from the linear relation between $\Delta R_{\textrm{NL4T}}$ and $\Delta R_{\textrm{ISHE}}$ (Eq. \ref{RisheNL4T}) to extract a large spin Hall angle $\theta_{\textrm{SH}}=(-19\pm4)\%$, which is simply ascribable to the sizable Berry curvature as aforementioned. This value is significantly larger than conventional FMs such as Py ($\sim1\%$ \cite{Tsukahara,SHEPy}), CoFe ($\sim5\%$ \cite{Wimmer}) and CoFeB ($\sim14\%$ \cite{Iihama}).

\begin{figure}
	\includegraphics[width=8.6cm]{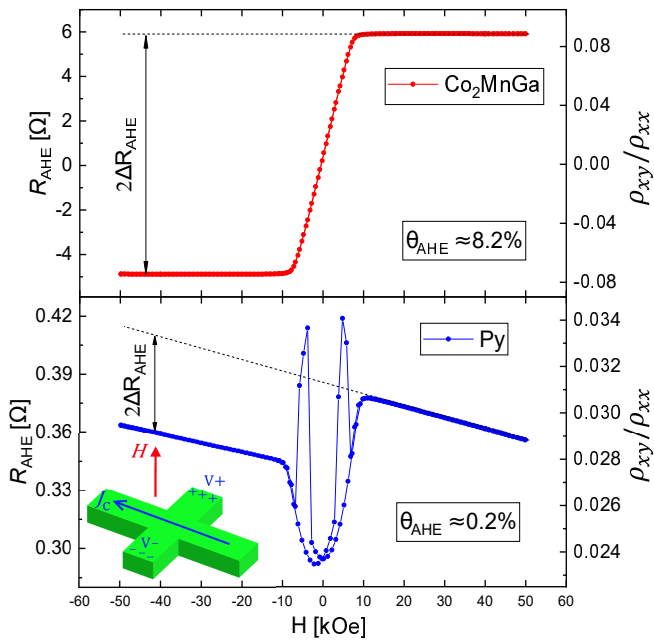}
	\centering
	\caption{Anomalous Hall resistance for Co$_2$MnGa (top panel) and Py (bottom panel) measured at room temperature. Inset of bottom panel shows a schematic of the measurement setup, where the magnetic field is applied out-of-plane in a Hall bar structure.}
	\label{Cinco}
\end{figure}

According to Onsager's reciprocal relation \cite{Onsager}, the resistance obtained for the SHE setup and ISHE setup should be the same, as it is the case for the Py and Pt control samples in Fig. \ref{Dos}(b). This was first experimentally demonstrated by Kimura et al. \cite{KimuraReversible} for the case of Pt and later verified for many other materials \cite{Niimi1,Niimi2,Oomori}. However, it was not the case for Co$_2$MnGa samples, where ISHE and SHE signals are clearly different, as shown in Fig. \ref{Dos}(c). It is important to note that there is no effect of the geometry of the devices on whether the reciprocity holds or not, as verified in the control experiments detailed in Sec. H of Supplemental Material \cite{SuppMat}.

While the origin of the reciprocity is well understood for nonmagnetic materials \cite{ReciprocalSHE}, it is still unclear why the relation should hold in FM systems, since there is a breaking of the time-reversal symmetry. Furthermore, recent advances in the field suggest that nonreciprocal transport can exist in chiral materials such as Weyl semimetals \cite{NonReciprocal}.

In case of Co$_2$MnGa, the non-reciprocity could be explained through a phenomenological picture of the Hall phenomena, by introducing the spin-dependent spin Hall angles $\theta_\uparrow=\sigma_{xy}^{\uparrow}/\sigma_{xx}^{\uparrow}$,  $\theta_\downarrow=\sigma_{xy}^{\downarrow}/\sigma_{xx}^{\downarrow}$ and the polarization of the spin Hall angle $p_{\textrm{SH}}$ through the relations $\theta_{\textrm{SH}}=(\theta_\uparrow+\theta_\downarrow)/2$ and $p_{\textrm{SH}}=(\theta_\uparrow-\theta_\downarrow)/(\theta_\uparrow+\theta_\downarrow)$, where $\sigma_{xy}^{s}$ $(s=\uparrow,\downarrow)$ is the spin-dependent Hall conductivity and $\sigma_{xx}^s$ $(s=\uparrow,\downarrow)$ is the spin-dependent normal conductivity \cite{Tung_2013,Oomori}. In the conventional case where $p_{\textrm{SH}}$ is assumed to be zero, the anomalous Hall angle $\theta_{\textrm{AHE}}$ is expected to be related to $\theta_{\textrm{SH}}$ via the spin polarization $\alpha_F$ in the form $\theta_{\textrm{AHE}}=\alpha_F\theta_{\textrm{SH}}$ \cite{Tsukahara}, as demonstrated for Py \cite{Oomori}. However, the value of $\theta_{\textrm{AHE}}$ observed for Co$_2$MnGa and Py have the same sign, as shown in Fig. \ref{Cinco}, while $\theta_{\textrm{SH}}$ exhibits different signs in both materials, as shown in Figs. \ref{Dos}(c) and \ref{Dos}(d). If a finite $p_{\textrm{SH}}$ is considered, the relation $\theta_{\textrm{AHE}}=(\alpha_F+p_{\textrm{SH}})\theta_{\textrm{SH}}$ should hold \cite{SuppMat}, indicating $p_{\textrm{SH}}$ should be negative and $|p_{\textrm{SH}}|>\alpha_F$. In addition, the relation between ISHE and SHE resistances obtained with the one dimensional spin diffusion model \cite{SuppMat}:

\begin{equation}
-\frac{\Delta R_{\textrm{ISHE}}}{\Delta R_{\textrm{SHE}}}=\frac{1-\alpha_F^2}{1+\alpha_F p_{\textrm{SH}}},
\end{equation}

\noindent indicates $|\Delta R_{\textrm{ISHE}}|>|\Delta R_{\textrm{SHE}}|$, which is clear in Fig. \ref{Dos}(c). Refer to Sec. G of Supplemental Material for a more detailed description of this spin-dependent spin Hall angle-based approach \cite{SuppMat}.

In conclusion, a direct and effective method to determine the spin Hall angle in ferromagnets was introduced, to obtain a significantly large value of $\theta_{\textrm{SH}}=(-19\pm4)\%$ for Heusler alloy Co$_2$MnGa, a Weyl semimetal. Combined with the ability to control the intensity of ISHE through the magnetization direction and the high resistivity of the compound \cite{SuppMat}, this result situates Co$_2$MnGa as a robust platform for detection and generation of spin currents in future spintronic devices. Furthermore, a lack of reciprocity between ISHE and SHE resistances was observed and attributed to a negative polarization of the spin Hall angle.
  
\hfill

This research was supported in part by a Grant-in-Aid
for Scientific Research from the Ministry of Education,
Culture, Sports, Science and Technology (MEXT) of Japan
(Innovative Area “Nano Spin Conversion
Science” KAKENHI No. 26103003), Grant-in-Aid for Young Scientists (A) No. 16H06089 and Grant-in-Aid for Scientific Research (S) "Semiconductor Spincurrentronics" No. 16H06330.
L. L. acknowledges support from MEXT doctoral scholarship. S. G. acknowledges financial support from the New Zealand Science for Technological Innovation National Science Challenge. The MacDiarmid Institute is supported under the New
Zealand Centres of Research Excellence Programme.

\bibliography{apssamp}

\begin{thebibliography}{58}
\bibitem [{Sup()}]{SuppMat}%
  \BibitemOpen
  \bibfield  {title} {\bibinfo {title} {See {S}upplemental {M}aterial at [url
  will be inserted by publisher] for details on sample preparation,
  determination of relevant parameters, characterization of magnetic
  electrodes, angle dependence of {I}{S}{H}{E} signal and discussion on the
  validity of reciprocal relations},\ }\href@noop {} {\ }\BibitemShut {NoStop}%
\end{thebibliography}

\pagebreak
\widetext
\begin{center}
	\textbf{\large Supplemental Materials: Giant spin Hall angle in the Heusler alloy Weyl ferromagnet Co\textsubscript{2}MnGa}
\end{center}
\setcounter{equation}{0}
\setcounter{figure}{0}
\setcounter{table}{0}
\makeatletter
\renewcommand{\theequation}{S\arabic{equation}}
\renewcommand{\thefigure}{S\arabic{figure}}

\subsection{Sample preparation}

Co$_2$MnGa films were grown using an ultra-high vacuum DC magnetron sputtering system with a base pressure of 2 $\times$10$^{-8}$ Torr at a temperature of 400 $^{\circ}$C, with a subsequent annealing at 550 $^{\circ}$C during 60 minutes. Since the growth of the films and the microfabrication process were made at  different places, a 40 nm-thick copper capping layer was grown on top of the Heusler alloy layer without breaking the vacuum but after cooling to room temperature. This capping layer was removed right before making the lateral spin valves. Injection and detection ferromagnetic electrodes ($F1$ and $F2$ in Fig. 1(b)) width varied between 150 nm and 200 nm. Different switching fields were achieved by changing the shape of the electrodes, favoring the domain rotation with a flag structure in case of $F1$, so the same width (at the interfaces with the nonmagnetic channel) could be used for both electrodes in a same device. The edge-to-edge separation, also known as channel length $L$, between the ferromagnetic electrodes varied from 150 nm to 1500 nm. In case of middle wire devices (Fig. 1(c) on main text) the width of the middle wire was between 60 nm and 150 nm, while $L$ was fixed at 600 nm. For Cu channels, a width $W_N$ of 200 nm and a thickness $t_N$ of 100 nm were used.

\subsection{Interface resistance measurement}

\begin{figure}[h]
	\includegraphics[width=12cm]{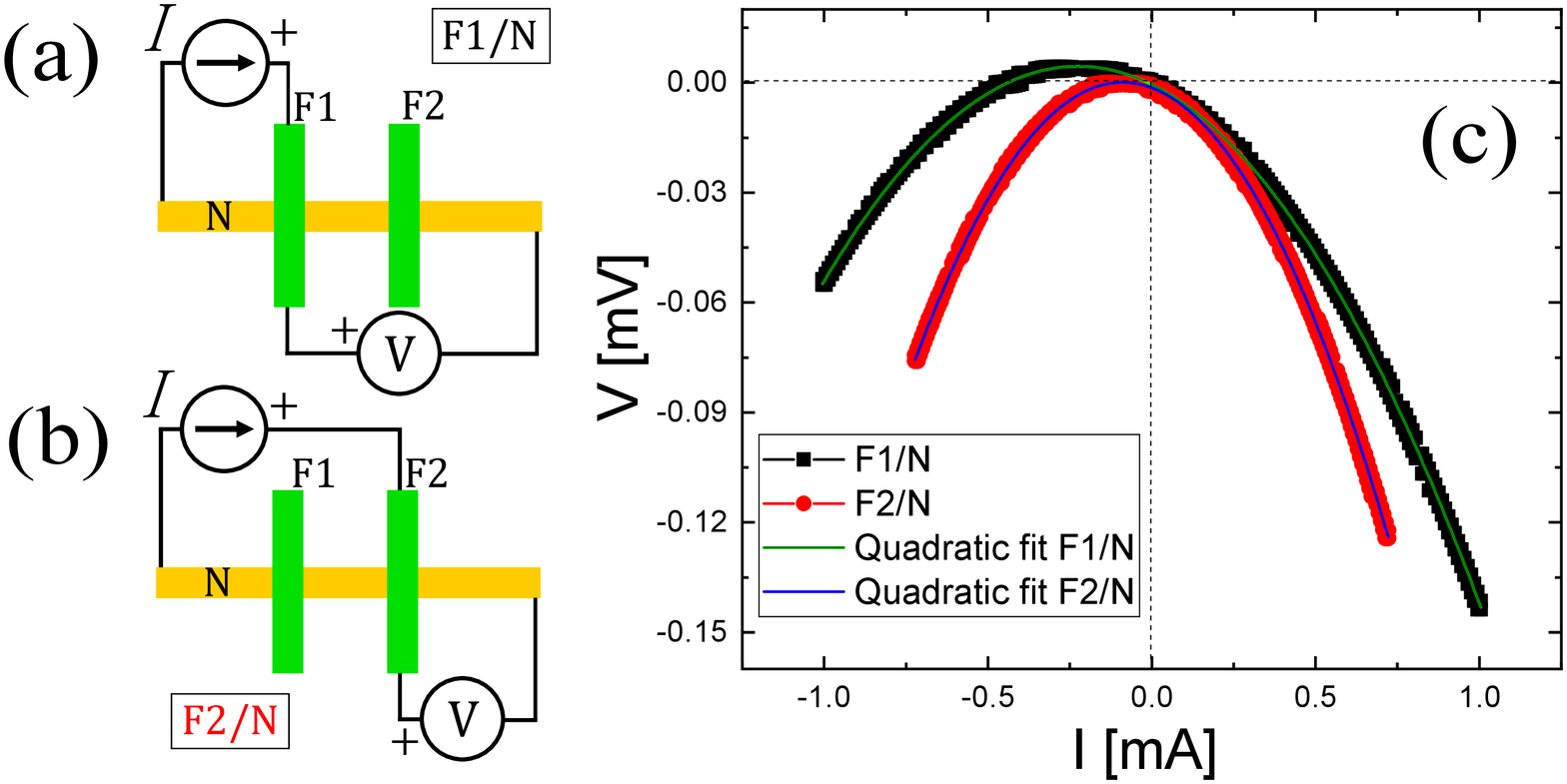}
	\centering
	\caption{Schematics of conventional cross configuration for interface resistance measurement for (a) the F1/N interface and (b) the F2/N interface. (c) Typical I-V curves where a quadratic fitting of the data was performed.}
	\label{intres}
\end{figure}

Obtaining transparent interfaces is important to reliably extract spin transport parameters such as spin polarization $\alpha_F$ and spin diffusion length $\lambda_F$ from the nonlocal resistance measurements. To this end, a 500V acceleration voltage Ar ion milling step is performed in-situ prior to Cu deposition. The interface resistance was measured at every device using the cross configuration shown in Fig. \ref{intres}, where typical I-V curves for the interfaces of $F1$ and $F2$ with $N$ are shown. A significant quadratic contribution was observed, attributed to thermally originated electromotive forces due to the temperature gradient in the interface \cite{Thermal}. Apart from that, a clear negative linear contribution of around -0.05 $\Omega$ is obtained from the polynomial fitting, as expected for a transparent interface. This measured negative resistance is an artifact that occurs when the resistance of the metal electrodes along the junction is similar or higher than the interface resistance of the junction itself \cite{Pedersen}. To estimate the real magnitude of the interface resistances, a finite element method simulation was performed, where, for a fixed current, the interface resistance was systematically changed until the voltage at the measurement points matched the experimental result. Calculation yield to a product of the interface resistance with the junction area $R_i A_{Fi}\lesssim2$ f$\Omega$m$^2$ in all reported cases.

\subsection{Determination of the spin diffusion length of absorption wire}

\begin{figure}
	\includegraphics[width=10cm]{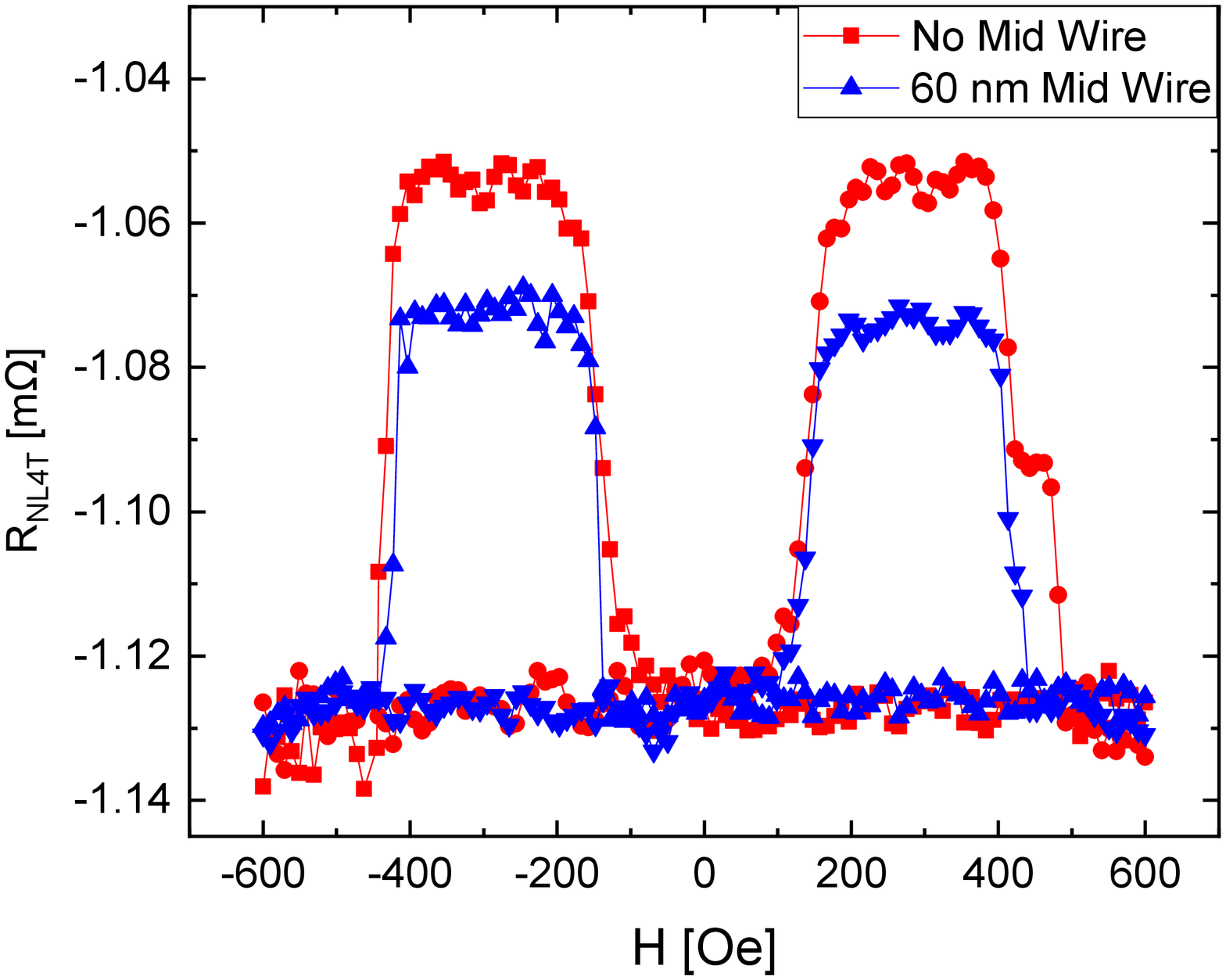}
	\centering
	\caption{Nonlocal 4 terminal resistances for Co$_2$MnGa LSVs as a function of a magnetic field applied parallel to FM electrodes, for a reference device with no middle wire and a another with a 60 nm middle wire.}
	\label{MidWire}
\end{figure}

To estimate the spin diffusion length of Co$_2$MnGa, the spin absorption technique was used. The experiment setup is equivalent to that for the nonlocal 4 terminal (NL4T) resistance measurement,  where $\Delta R_{NL4T}$ is obtained, as detailed in the main text. Electrodes $F1$ and $F2$ were used as injector and detector, respectively (see Fig. 1(c)), in devices with different widths of the middle wire $M$, obtaining $\Delta R_{\textrm{NL4T}}^M$. Additionally, devices with the same separation between $F1$ and $F2$ but no middle wire were used as reference samples, obtaining $\Delta R_{\textrm{NL4T}}^{ref}$. The signal is expected to decrease with the presence of the middle wire, since part of the spin current is absorbed there. The quotient between these two signals can be calculated using the 1D spin diffusion model, leading to:

\begin{equation}
\frac{\Delta R_{\textrm{NL4T}}^M}{\Delta R_{\textrm{NL4T}}^{ref}}=\frac{2R_M(R_N+e^{L/\lambda_N}(2R_{F}+R_N))}{(2R_M-R_N)R_N+e^{L/\lambda_N}(2R_F+R_N)(2R_M+R_N)},
\tag{S1}
\label{S1}
\end{equation}

\noindent where it was assumed $R_{F1}=R_{F2}=R_F$ and the position of the middle wire $M$, in a point contact model, is right at the middle of the gap. The spin resistance $R_X$ is defined as $\lambda_X\rho_X/(1-\alpha_X^2)A_X$, where $\lambda_X$, $\rho_X$, $\alpha_X$ and $A_X$ denotes the electrical resistivity, the spin diffusion length, the spin polarization and the effective cross sectional area, respectively, for an specific ``X" material. A typical measurement is shown in Fig. \ref{MidWire}, where $M$ is the same material as $F1$ and $F2$, and thus it has the same spin polarization $\alpha_F$ and spin diffusion length $\lambda_F$. In order to determine $\lambda_F$, equation \ref{S1} should be used together with Eq. (5) and the data of Fig. 3 of main text, where the $L$ dependence of $\Delta R_{\textrm{NL4T}}$ for devices with no middle wire was studied. Unknown parameters are $\lambda_N$, $\lambda_F$ and $\alpha_F$, since the resistivities were determined with a 4-probe method obtaining $\rho_{\textrm{Cu}}=(2.8\pm0.1)$ $\mu \Omega$cm and $\rho_{\textrm{Co$_2$MnGa}}=(220\pm30)$ $\mu \Omega$cm, which agree with reports in the literature \cite{RhoN,RhoF}, and the effective cross sectional areas depend on the sample geometry that was determined with Atomic Force Microscopy technique and SEM imaging. By iterating between equations \ref{S1} and 5, the three unknown parameters where extracted consistently obtaining $\alpha_F=0.15\pm0.03$, $\lambda_N=(410\pm80)$ nm and $\lambda_F=(3.1\pm1.1)$ nm. This latter value is in the order of the reported spin diffusion lengths for similar Heusler compounds \cite{LFHeusler}.

\subsection{Determination of the shunting factor $x$}

\begin{figure}
	\includegraphics[width=16cm]{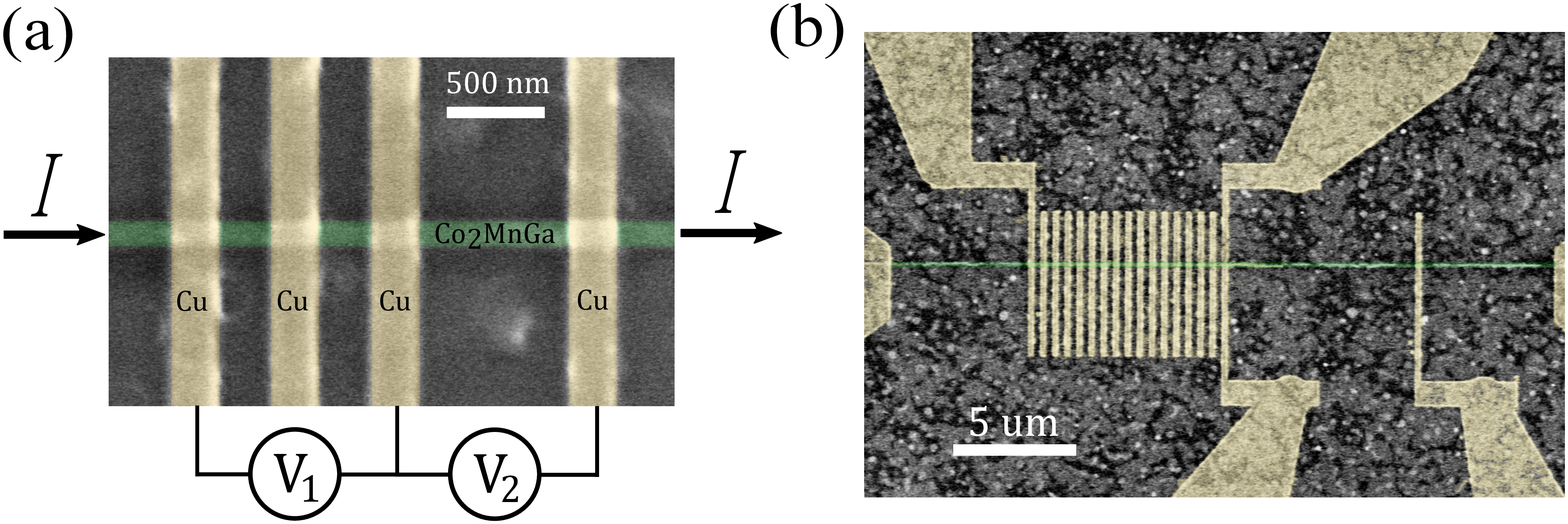}
	\centering
	\caption{(a) SEM image of a device with a single Cu shunting wire used for the estimation of the shunting factor, with schematics of the voltmeters connections. (b) SEM image of a similar device used for the estimation of the shunting factor but with several (15) Cu shunting wires.}
	\label{Shunting}
\end{figure}

In the ISHE setup, a spin current is absorbed by the ferromagnet F2 from the nonmagnetic channel N. Since $\lambda_F$ is, as previously calculated, in the order of a few nm, the spin to charge conversion will occur mostly very close to the N/F2 interface. Therefore, it is expected that some of the ISHE-generated charge current at the $FM$ will shunt through the much more conductive Cu channel. This will produce a decrease of the generated voltage, in a fraction $x$ ($0<x<1$) of the total expected voltage if no shunting existed. This can also be understood as $x$ being the fraction of the total current that will go through the FM. In a similar way, in the SHE setup, a charge current $I_c$ is injected at the FM, and it is also expected that, at the N/F2 interface, part of the current will shunt through the Cu wire meanwhile a fraction $x$ will remain in the FM. This will produce a decrease of the voltage drop across the FM wire, compared to when no shunting happens. In order to estimate the magnitude of this effect, that is quite critical for the spin Hall angle calculation, multi-terminal devices such as those of Fig. \ref{Shunting} were fabricated. The experiment is described in Fig. \ref{Shunting}(a), where a charge current is injected through the Co$_2$MnGa electrode terminals. The voltage $V_1$ is expected to be less than $V_2$ due to the additional shunting through the Cu middle wire, following the relation:

\begin{equation}
\frac{V1}{V2}=\frac{2W_2+2W_1 x}{2W_2+W_1(x+1)},
\tag{S2}
\end{equation}

\noindent where $W_1$ is the width of the Cu wires and $W_2$ is the separation between adjacent Cu wires.

Since there was a significant dispersion on the $x$ values extracted by this method, due to process variations, devices with several shunting middle wires were designed, as shown in Fig. \ref{Shunting}(b), in order to average the shunting effect through the multiple wires. The widths of the Cu and Co$_2$MnGa wires were selected to replicate those of the lateral spin valves used for the SHE measurements. Finally, the reciprocal of the shunting factor was found to be $x^{-1}=(2.63\pm0.37)$.

\subsection{Angular dependence of inverse spin Hall effect voltage}

\begin{figure}
	\includegraphics[width=14cm]{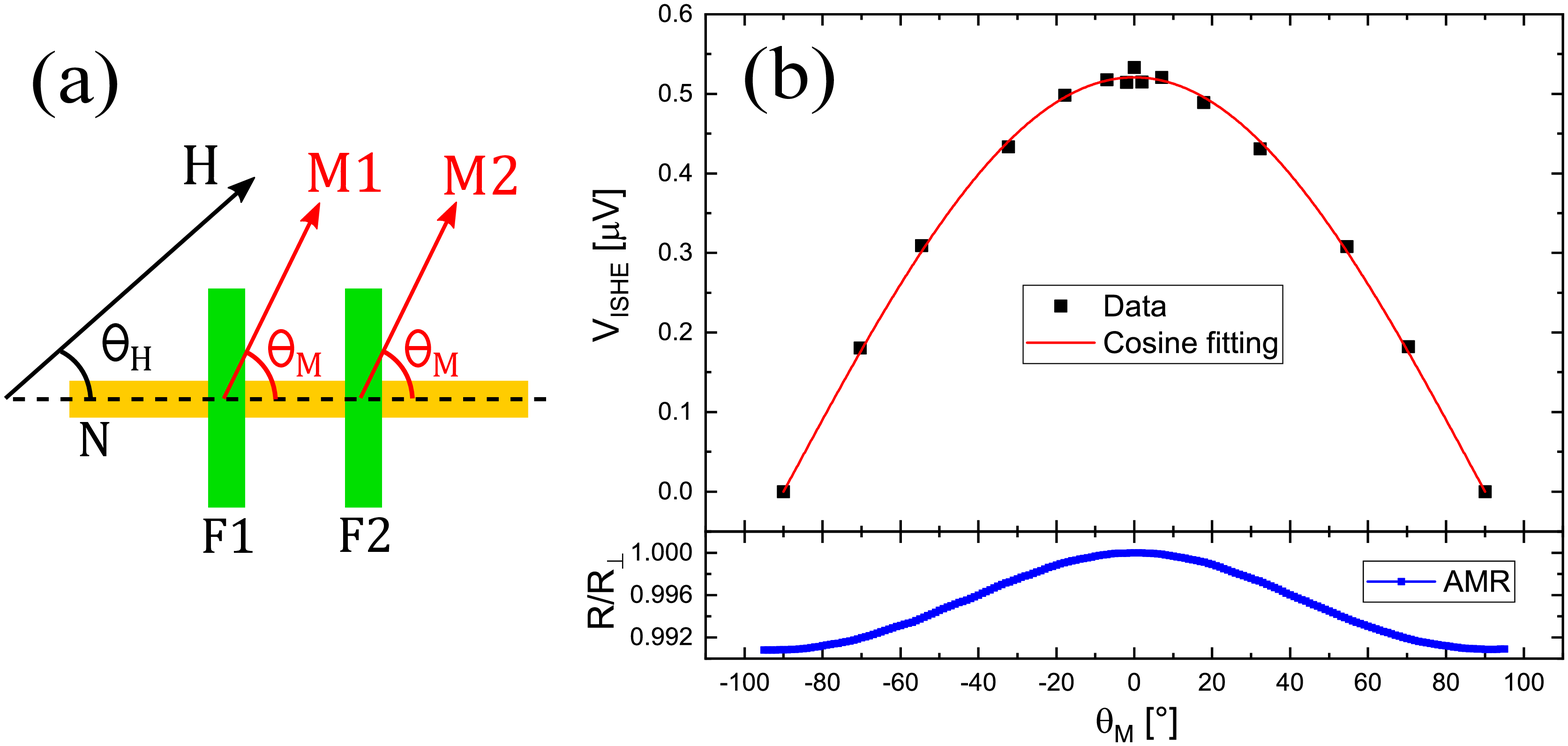}
	\centering
	\caption{(a) Schematics of the geometry for applying the external magnetic field $H$. To note is that in principle $\theta_H\neq\theta_M$. (b) Magnetization angle dependence of ISHE voltage. Bottom panel shows the AMR curve obtained with an external magnetic field of 60 kOe.
	}
	\label{Angle}
\end{figure}

In order to prove that the magnitude of the inverse spin Hall effect signals can be controlled through the magnetization orientation, the dependence of the ISHE voltage with the angle of the magnetization was studied. To this end, the angle of the external magnetic field $\theta_H$ was varied, as shown in Fig. \ref{Angle}(a). Since a magnetic anisotropy exists due to the electrode shape, $\theta_H$ in principle will not coincide with the angle of the magnetization $\theta_M$. AMR measurements for $F1$ and $F2$ confirmed that both magnetizations $M1$ and $M2$ remain approximately parallel to each other as they rotates due to the rotation of the external field. Since the maximum value used for the sweep of the magnetic field in the ISHE measurements ($\sim$4 kOe) was not enough to assure $H$ and $M$ were parallel, a correction was made to convert $\theta_H$ to the real $\theta_M$. To this end, the AMR curve of $F1$ was measured for a sufficiently high magnetic field of 60 kOe so $\theta_H=\theta_M$, shown in Fig. \ref{Angle}(b). The magnetic field angle dependence of the resistance was also obtained with a field of 4kOe to determine $\theta_M$ by matching the resistances with the high field curve. The ISHE voltage, for a same injection current of 1 mA, as a function of $\theta_M$ is shown in Fig. \ref{Angle}(b), where the expected cosine behavior confirms the signal magnitude is being controlled through the magnetization direction.

\subsection{Characterization of ferromagnetic electrodes}

\begin{figure}
	\includegraphics[width=17cm]{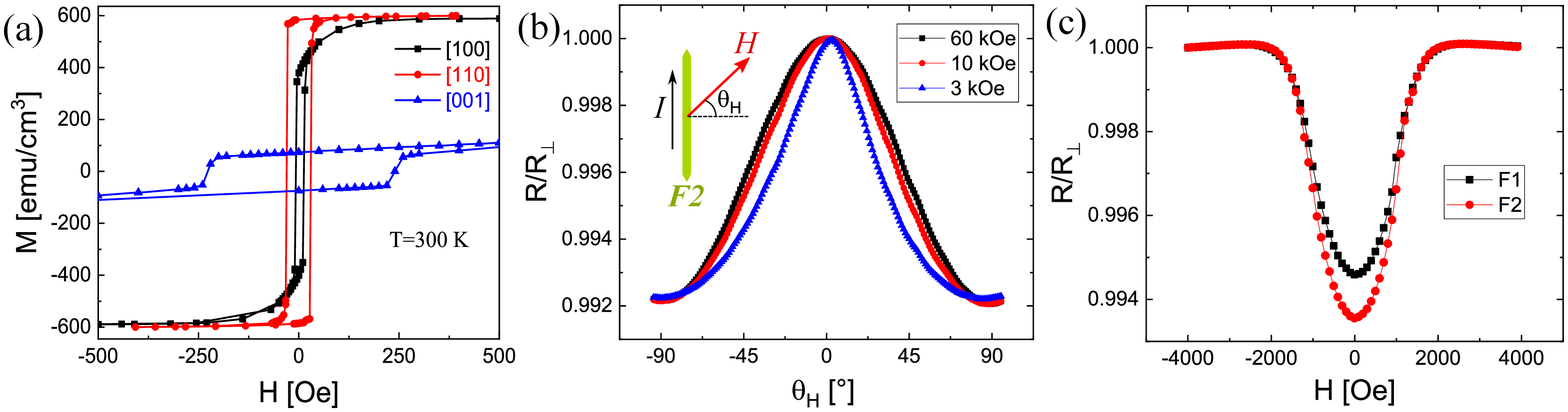}
	\centering
	\caption{(a) Magnetization hysteresis loop of a Co$_2$MnGa film measured for three different crystalline directions at room temperature. (b) AMR of a Co$_2$MnGa ferromagnetic electrode, F2 in this case, as a function of the angle of the applied magnetic field in the plane of the substrate, for three different magnetic fields. (c) AMR of both ferromagnetic electrodes of a Co$_2$MnGa LSV as a function of the intensity of magnetic field for a fixed angle $\theta=0^\circ$, this is, the same condition of ISHE/SHE experiments.
	}
	\label{Mag}
\end{figure}

Before patterning of ferromagnetic electrodes, the magnetization of Co$_2$MnGa thin films was measured using a SQUID magnetometer along the crystalline directions [100], [110] and [001]. As shown in the XRD pattern of Fig. 1(a) in the main text, the [001] axis is out of the substrate plane, and therefore the saturation occurs at a high magnetic field of 10 kOe (not shown) due to shape anisotropy. Concerning the in-plane directions, it was found the existence of magneto-crystalline anisotropy with easy axis along the [110] direction (Fig. \ref{Mag}(a)), which was verified to be parallel to the substrate sides. However, after the patterning of FM electrodes, the shape anisotropy of these thin wires was found to be much more significant than the original magneto-crystalline anisotropy of the films. In Fig. \ref{Mag}(b), it can be seen that for a high enough magnetic field of 60 kOe, magnetization and magnetic field are parallel, showing the AMR a $\cos^2\theta$ behavior. However, for a smaller magnetic field of 3 kOe, it is clear how the shape anisotropy makes the magnetization to be closer to the direction parallel to electrode length. When the in-plane magnetic field direction is fixed perpendicular to ferromagnetic electrodes length ($\theta=0^\circ$), this is, the same condition as in SHE and ISHE measurements, it can be seen that the AMR of both F1 and F2 electrodes saturates for a field around 2000 Oe. The relative amplitude of AMR is smaller in F1 electrode, but this is due to its different shape, as shown in Fig. 1(b) in the main text, where part of the electric current is not parallel to the electrode length. However, the AMR curves for both F1 and F2 are the same except for a scale factor. The conclusion is that the magnetization of both ferromagnetic electrodes rotate coherently when the magnetic field is swept, probably because F1 and F2 have the same width. 

\subsection{On the validity of Onsager's reciprocal relation}

In order to explain the origin of the breaking of reciprocity observed in the ISHE and SHE experiments on Co$_2$MnGa, a phenomenological picture is presented. Assuming the validity of the two channel model, the total charge current density $\vect{j_c}$ is decomposed into the spin components as $\vect{j_c}=\vect{j_\uparrow}+\vect{j_\downarrow}$, meanwhile the spin current (with charge current units) is defined as $\vect{j_s}=\vect{j_\uparrow}-\vect{j_\downarrow}$. The spin Hall effect-generated currents $\vect{j_\uparrow}^{H}$ and $\vect{j_\downarrow}^H$ can be described for each spin-component of the current as $\vect{j_\uparrow}^{H}=\theta_\uparrow \vect{\hat s}\times\vect{j_\uparrow}$ and $\vect{j_\downarrow}^{H}=-\theta_\downarrow \vect{\hat s}\times\vect{j_\downarrow}$, where $\theta_\uparrow=\sigma_{xy}^{\uparrow}/\sigma_{xx}^{\uparrow}$ and $\theta_\downarrow=\sigma_{xy}^{\downarrow}/\sigma_{xx}^{\downarrow}$ are the spin-dependent spin Hall angles, defined in terms of the spin-dependent Hall $\sigma_{xy}^{s}$ $(s=\uparrow,\downarrow)$ and normal $\sigma_{xx}^s$ $(s=\uparrow,\downarrow)$ conductivities. The spin Hall angle is defined as $\theta_{\textrm{SH}}=(\theta_\uparrow+\theta_\downarrow)/2$ and the polarization of the spin Hall angle is introduced as $p_{\textrm{SH}}=(\theta_\uparrow-\theta_\downarrow)/(\theta_\uparrow+\theta_\downarrow)$.

\begin{figure}
	\includegraphics[width=12cm]{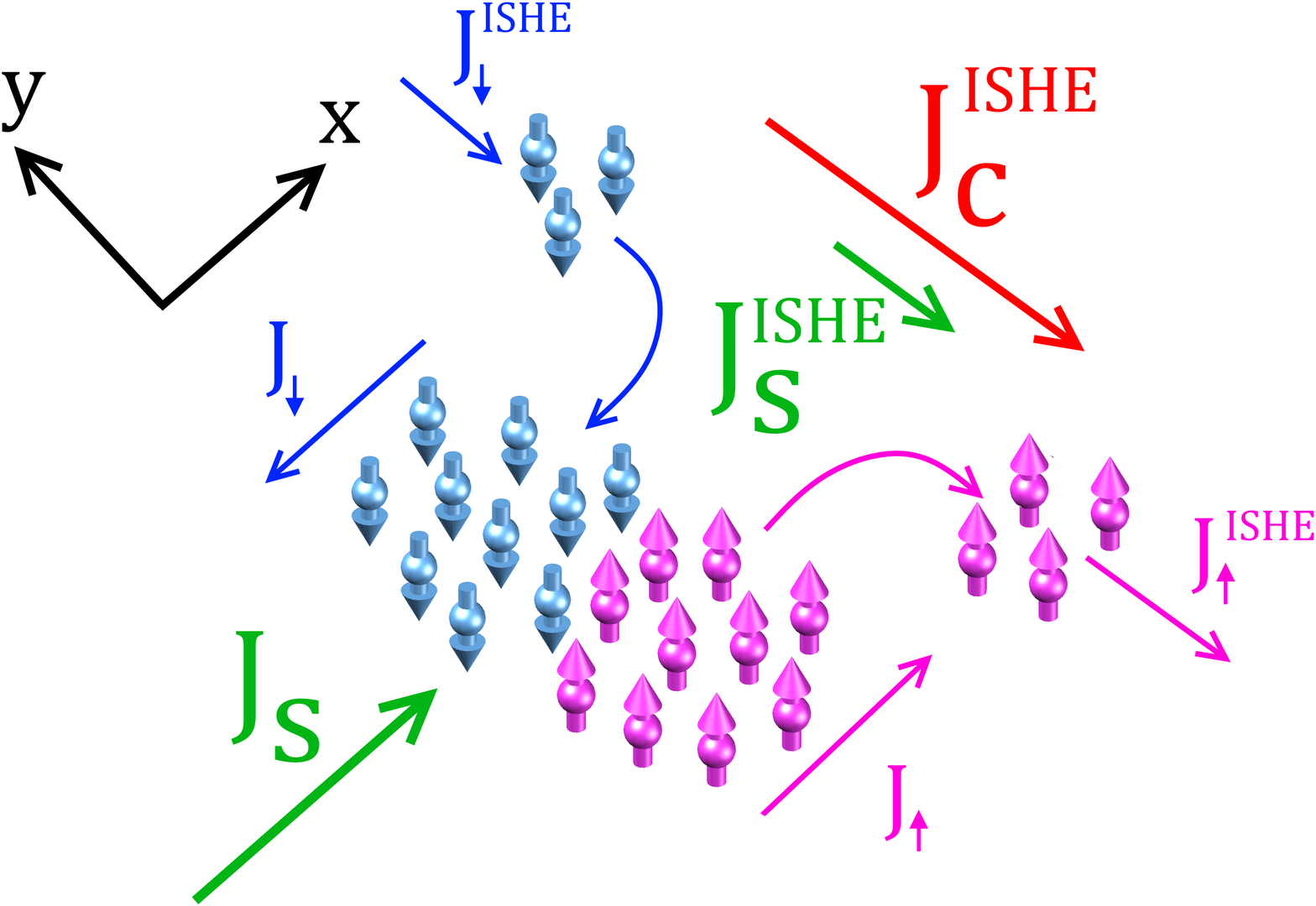}
	\centering
	\caption{Schematics of electron motion when a pure spin current induces a spin polarized charge current via the spin dependent ISHE.}
	\label{ISHE}
\end{figure}

The ISHE case is shown schematically in Fig. \ref{ISHE}, where the direction of electron motion is that of the charge flow, for simplicity. In this kind of experiment, the source is a diffusive pure spin current $\vect{j_s}$ that is absorbed by the FM material, so $j_\uparrow=-j_\downarrow=j_s/2$ along the $x$ direction. Then, the ISHE-generated charge current along $y$ direction is $j_c^{\textrm{ISHE}}=\theta_{\textrm{SH}}j_s$, meanwhile the generated spin current is $j_s^{\textrm{ISHE}}=\theta_{\textrm{SH}}p_{SH}j_s$. Since the measured ISHE voltage is proportional to $j_c^{\textrm{ISHE}}$, this method allows to extract $\theta_{\textrm{SH}}$ regardless of the value of $p_{SH}$.

\begin{figure}
	\includegraphics[width=12cm]{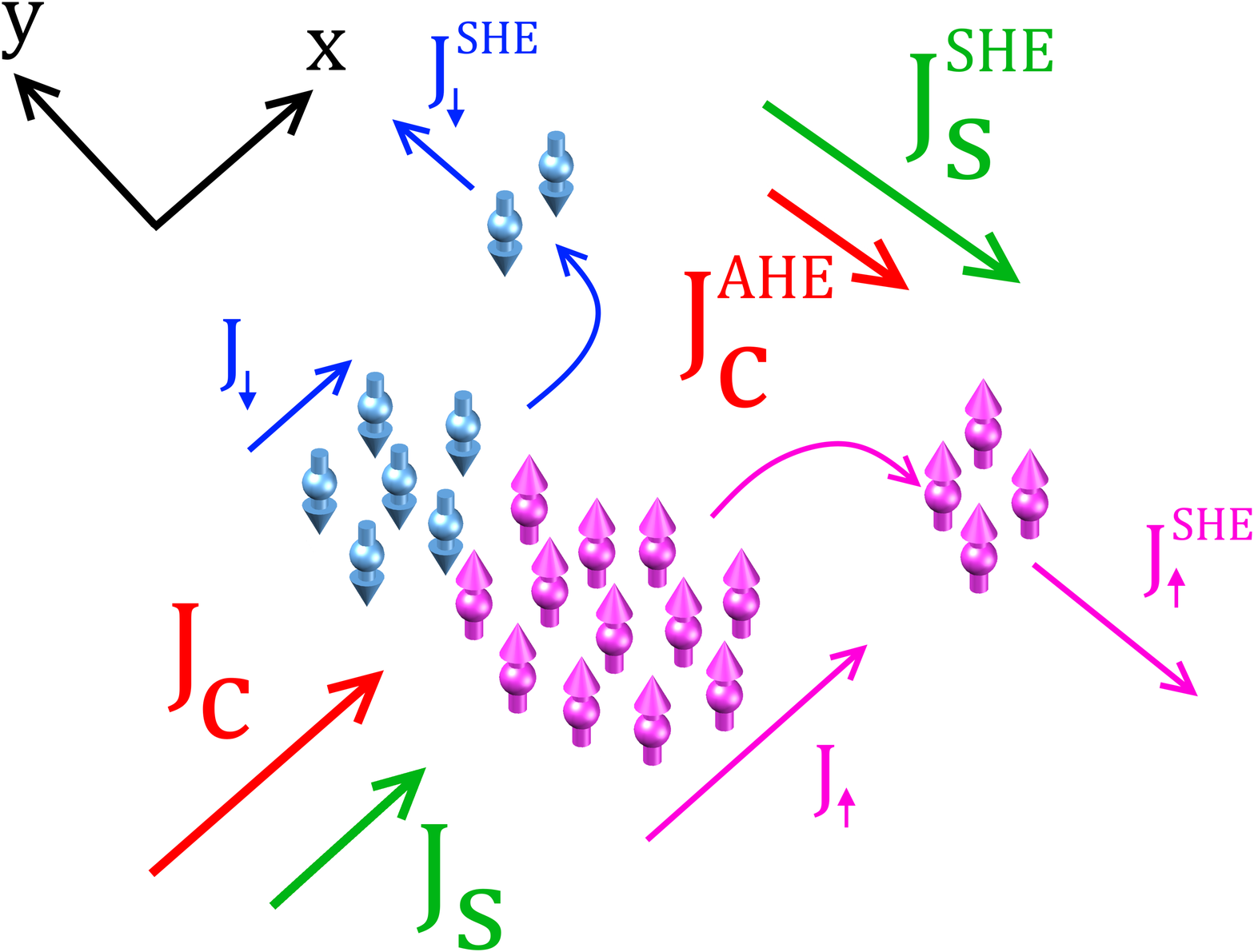}
	\centering
	\caption{Schematics of electron motion when a spin polarized charge current induces both a charge current, known as anomalous Hall effect, and a spin current via the spin dependent SHE.}
	\label{SHE}
\end{figure}

In contrast, the SHE case, shown schematically in Fig. \ref{SHE}, is essentially different since the source is not a pure charge current (as it would be in a nonmagnetic material) but a spin polarized charge current, because of the spin polarization $\alpha_F$ of the FM, defined as $\alpha_F=(\sigma_\uparrow-\sigma_\downarrow)/(\sigma_\uparrow+\sigma_\downarrow)$. Therefore the spin components of the current along $x$ direction are $j_\uparrow=j_c(1+\alpha_F)/2$ and $j_\downarrow=j_c(1-\alpha_F)/2$. Then, the SHE-generated charge current along the $y$ direction, also known as anomalous Hall effect current, is $j_c^{\textrm{AHE}}=\theta_{\textrm{SH}}(\alpha_F+p_{SH})j_c$ and the anomalous spin Hall angle $\theta_{\textrm{AHE}}$ can be expressed as $\theta_{\textrm{AHE}}=\theta_{\textrm{SH}}(\alpha_F+p_{\textrm{SH}})$, where it is clear that $\theta_{\textrm{AHE}}$ and $\theta_{\textrm{SH}}$ may have different sign if $p_{SH}$ is negative. In Fig. \ref{SHE} diagram, it would mean a larger number of spin-down (blue) electrons will scatter to left. If this number is high enough, the direction of $j_c^{\textrm{AHE}}$ can be inverted while the direction of $j_s^{\textrm{SHE}}$ remains the same. On the other hand, the SHE-generated spin current is $j_s^{\textrm{SHE}}=\theta_{\textrm{SH}}(1+\alpha_Fp_{SH})j_c$, where a non-zero $p_{SH}$ would imply an asymmetry with respect to the ISHE case.

To calculate the SHE nonlocal resistance measured in the SHE setup for a 2-wire device, as described in the main text, a similar derivation as used for the NL4T resistance (Eq. 5 on main text) can be done. While the detection mechanism is essentially the same, the spin injection is different in SHE and NL4T cases. In electrical spin injection, the source of spin accumulation is an electric-field-driven drift spin current in the ferromagnet, while in SHE-driven spin injection, the source of spin accumulation is also a drift spin current but generated via the SHE. Then, the boundary conditions are modified at the injection interface. This imply that one of the $\alpha_F$ factors in Eq. 5 of main text ends up being replaced by  $\theta_{\textrm{SH}}(1+\alpha_Fp_{SH})xW_N/t_F$, being the final expression for $\Delta R_{\textrm{SHE}}$: 

\begin{equation}
\Delta R_{\textrm{SHE}}=\frac{\theta_{SH} x W_N}{t_{F}}\times\frac{(1+\alpha_Fp_{SH})2\alpha_{F}R_F^2R_Ne^{-L/\lambda_N}}{(2R_{F}+R_N)^2-R_N^2e^{-2L/\lambda_N}}.
\label{RSHE}
\tag{S3}
\end{equation}

By comparing Eq. \ref{RSHE} with the expression for $\Delta R_{\textrm{ISHE}}$ of Eq. 2 in main text, the Eq. 6 of the main manuscript is obtained, where the lack of reciprocity is exposed. This model consistently explains that a different sign for $\theta_{SH}$ and $\theta_{\textrm{AHE}}$ implies the SHE resistance is reduced with respect to the ISHE resistance, which agrees with the experimental results in Fig. 2(c) of main text.

It should be pointed out that the above decomposition of the spin Hall effect in two spin-conserving effects for up-spin and down-spin carriers is not a fully precise picture, since it is known the spin-orbit coupling generally mixes spin-up and spin-down states through spin flip processes. However, for light elements such as 3d transition metals, the perfect two-current model is expected to be still reasonably valid \cite{Zhang}.

\subsection{Checking the influence of sample geometry on the reciprocity of transport signals}

\begin{figure}[H]
	\includegraphics[width=15cm]{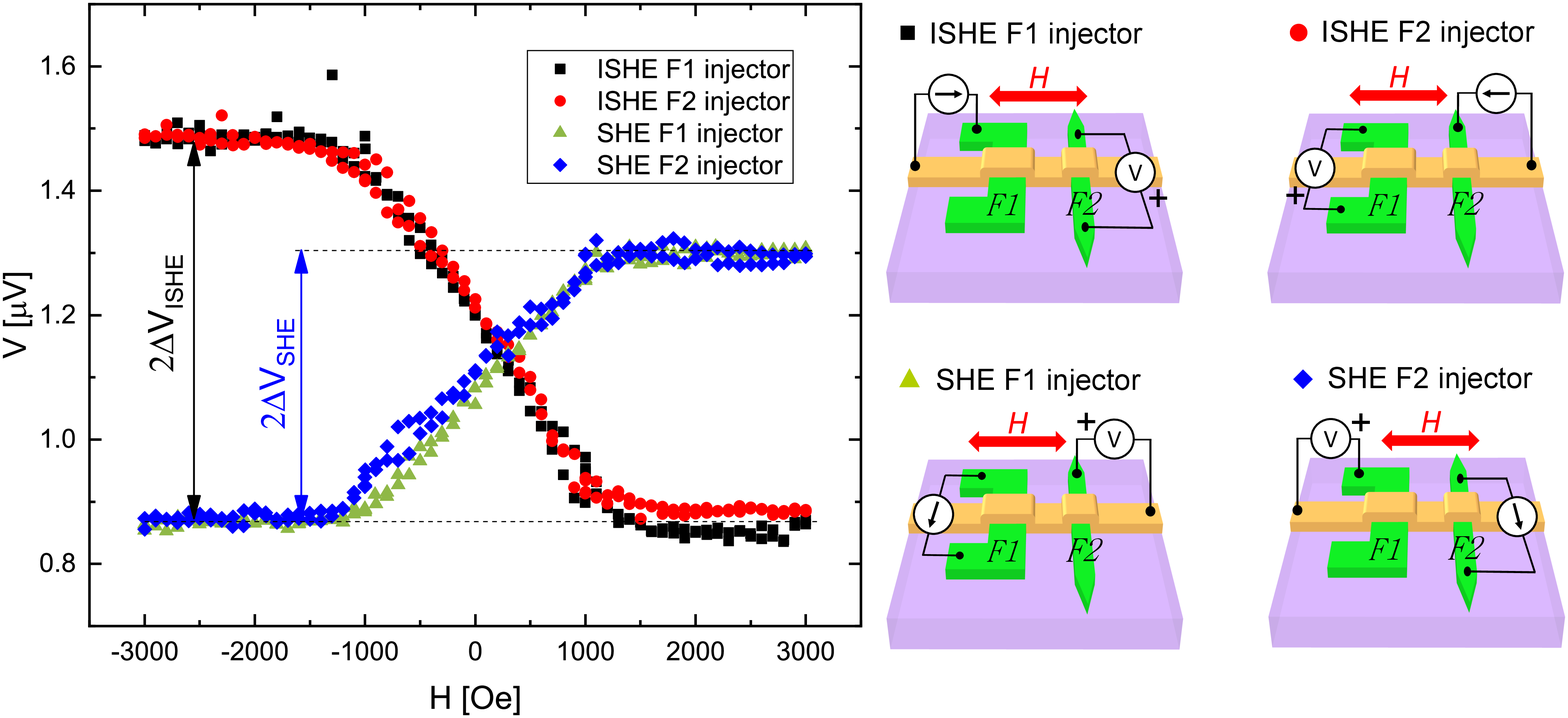}
	\centering
	\caption{ISHE and SHE signals measured for an injection current of 1mA, at several different configurations changing the injection and detection electrodes, detailed in the schematics.}
	\label{S7}
\end{figure}

As shown in the SEM image of Fig. 1(b) of the main text, as well as in the schematics of Fig. \ref{S7}, the ferromagnetic electrodes F1 and F2 have different shape. This design is in order to achieve different switching fields when the magnetic field is parallel to the electrode length in the nonlocal-4-terminal setup. Therefore, F1 and F2 are not equivalent, and the measured signals when using F1 as injector and F2 as detector and vice versa were checked, both for ISHE and SHE setups, as shown in Fig. \ref{S7}. It can be seen that, regardless of which electrodes are used for injection and detection, the measured amplitudes are the same, either for the SHE or the ISHE setup. Thus, any possible influence of the non-symmetrical geometry on the non-reciprocity is discarded. At the same time, the breaking of the reciprocity, this is, different amplitudes for $\Delta V_{\textrm{ISHE}}$ and $\Delta V_{\textrm{SHE}}$ (for the same injection current), were consistently observed again in random device tests made for this check.

\end{document}